\documentclass[preprint,prb,amssymb,showpacs]{revtex4}
\usepackage{graphicx}
\usepackage{dcolumn}
\usepackage{amsmath}

\begin{document}
\title{Transverse negative
magnetoresistance of 2D structures in the presence of strong
in-plane magnetic field: weak localization as a probe of interface
roughness}
\author{G.~M.~Minkov}
\email{Grigori.Minkov@usu.ru} \author{O.~E.~Rut}
\author{A.~V.~Germanenko}
\author{A.~A.~Sherstobitov}
\affiliation{Institute of Physics and Applied Mathematics, Ural
State University, 620083 Ekaterinburg, Russia}

\author{B.~N.~Zvonkov}
\affiliation{Physical-Technical Research Institute, University of
Nizhni Novgorod, Nizhni Novgorod 603600, Russia}

\author{D.~O.~Filatov}
\affiliation {Research and Educational Center for Physics of the
Solid State Nanostructures, University of Nizhny Novgorod, Nizhni
Novgorod 603600, Russia}
\date{\today}

\begin{abstract}
The interference induced transverse negative magnetoresistance of
GaAs/In$_{x}$Ga$_{1-x}$As/GaAs quantum well heterostructures has
been studied in the presence of strong in-plane magnetic field. It
is shown that effect of in-plane magnetic field is determined by
the interface roughness and strongly depends on the relationship
between mean free path, phase breaking length and roughness
correlation length. Analysis of the experimental results allows us
to estimate parameters of short- and long-range correlated
roughness which have been found in a good agreement with atomic
force microscopy data obtained for just the same samples.
\end{abstract}

\maketitle

\section{Introduction}
\label{sec:intr}

The interference correction determines in the main the temperature
and magnetic field dependences of the conductivity of weakly
disordered  two-dimensional (2D) systems. This correction
originates in the constructive interference of time-reversed
electron trajectories.  For an ideal two-dimensional gas of
spin-less particles only perpendicular magnetic field $B_\perp$
destroys the interference resulting in negative magnetoresistance,
whereas an in-plane magnetic field $B_\parallel$ does not effect
the interference correction. \cite{Altshuler} Therefore, an
applying of  in-plane magnetic field should not change the
magnetoresistance caused by the perpendicular field.

For real 2D system placed in in-plane magnetic field an interface
roughness leads to that an electron effectively feels random
perpendicular magnetic field at motion. This effect not only gives
rise to the negative magnetoresistance at in-plane magnetic field
but changes the magnetoresistance in perpendicular field in the
presence of in-plane field.

Another mechanism of influence of in-plane magnetic field on the
interference correction is the spin relaxation. If the relaxation
is strong enough it suppresses the weak localization leading to
positive magnetoresistance in very low magnetic field and
appearance of so-called antilocalization maximum on the
resistivity-magnetic field curve. In-plane magnetic field
resulting in Zeeman splitting decreases the spin relaxation rate
and, thus, affects the interference induced magnetoresistance.
\cite{Malshukov0} This mechanism is not effective in considerably
dirty systems, in which the spin-relaxation rate is less than the
dephasing rate and antilocalization maximum is not evident even at
very low temperature.

Anomalous low field magnetoresistance at perpendicular magnetic
field  has been studied extensively for many years. Considerably
less attention has been directed to the effect of in-plane
magnetic field  on interference correction. Although the effects
are small they are interesting because give an information on
interface roughness of 2D structures. The first detailed
experimental study of effects of in-plane magnetic field on
negative magnetoresistance at perpendicular field was carried out
in  Ref.~\onlinecite{Wheeler} for silicon MOSFETs. There was shown
that short-range correlated roughness ($L<l_p$, where $L$ is
distance over which fluctuations are correlated and $l_p$ is mean
free path) leads to decreasing of phase breaking time
($\tau_\varphi$) by in-plane magnetic field and thus to changing
of the shape of magnetoresistance curve. The  theoretical analysis
carried out in recent paper by Mathur and Baranger\cite{Mathur}
shows that effect of in-plane magnetic field on the shape of
magnetoresistance curve strongly depends on relationship between
$L$, $l_p$ and $l_\varphi=\sqrt{D \tau_\varphi}$, where $D$ is
diffusion coefficient. Thus, the experimental studies of the
interference correction in the presence of in-plane magnetic field
gives a possibility to find the parameters of interface roughness
in particular structure.

This paper is devoted to the experimental study of the
interference induced transverse negative magnetoresistance of
GaAs/In$_{x}$Ga$_{1-x}$As/GaAs quantum wells with different scales
of interface roughness in the presence of strong in-plane magnetic
field. It is organized as follows. In the next section we give
experimental details. Experimental results are presented and
discussed in Sec.~\ref{sec:resultsWL}: Sec.~\ref{ssec:3512} and
Sec.~\ref{ssec:H5610} are dedicated to results of
magnetoresistance measurements for the samples with short- and
long-correlated roughness, respectively, Sec.~\ref{ssec:AFM} is
concerned with the results of atomic force microscopy.

\section{Experimental details}
\label{sec:ed}

In the present work we experimentally study the two types of
single quantum well heterostructures. The structure 3512 is
GaAs/InGaAs/GaAs quantum well heterostructure which consists of
0.5 mkm-thick undoped GaAs epilayer, a Sn $\delta$-layer, a 9 nm
spacer of undoped GaAs, a 8 nm In$_{0.2}$Ga$_{0.8}$As well, a 9 nm
spacer of undoped GaAs, a Sn $\delta$-layer, and a 300 nm cap
layer of undoped GaAs. In the second structure, H5610, the
arrangement of the doped layers was the same as in the first one.
The difference only is that the thin layer of InAs instead of
In$_{0.2}$Ga$_{0.8}$As layer has been grown. The large lattice
mismatch between InAs and GaAs results in this case in formation
of nanoclusters. They are situated on the InAs wetting layer of
one-two monolayers thickness, which is thin deep quantum well for
electrons. The samples were mesa etched into standard Hall bars
and then an Al gate electrode was deposited by thermal evaporation
onto the cap layer through a mask. Applying the gate voltage $V_g$
we were able to change the electron density and conductivity of 2D
gas. At electron density higher than approximately $7\times
10^{11}$ cm$^{-2}$ for structure 3512 and $9\times 10^{11}$
cm$^{-2}$ for structure H5610, the states in $\delta$-layers start
to be occupied that affects the dephasing rate and influences the
magnetoresistance curve. \cite{our1} In the present paper we
restrict our consideration by the case when the states in
$\delta$-layers are empty. In opposite case additional effects in
parallel magnetic field occur which will be considered elsewhere.
The structures parameters for some gate voltage are presented in
Table~\ref{tab1}.

In order to apply tesla-scale in-plane magnetic field while
sweeping subgauss control of perpendicular field, we mount the
sample with 2D electrons aligned to the axis of primary solenoid
(accurate to $\sim 1^\circ$) and use an independent split-coil
solenoid  to provide $B_\perp$ as well as to compensate for sample
misalignment. The two calibrated Hall probes were used to measure
$B_\perp$ and $B_\parallel$.

\begin{table}[tbp]
\caption{The parameters for the structures for different gate
voltages} \label{tab1}
\begin{ruledtabular}
\begin{tabular}{ccccccc}
Structure & $V_g$ (V) & n(10$^{12}$ cm$^{-2}$) &
$\sigma(G_0)$\footnotemark[1]
& $\sigma_0(G_0)$\footnotemark[2]&$\tau_p(10^{-13}\,\text{s})$ & $B_{tr}(\text{T})$\\
 \colrule
 3512  & 0.0   &1.0    &165.08 &169.5  &4.4    &0.0071\\
                &-0.5   &0.88   &122.95 &127.6  &3.8    &0.011\\
                &-0.75  &0.69   &83.61  & 88.7  &3.4    &0.018\\
                &-1.0   &0.67   &70.4   &75.5   &3.0    &0.024\\
                &-1.5   &0.47   &20.35  &26.4   &1.47    &0.138\\
                &-2.5   &0.32   &4.27   &9.3    &0.76   &0.76\\
\colrule
5610\#1\footnotemark[3]       &-1.0   &0.91   &38.8   &45.3  &1.31    &0.091\\
                &-2.5   &0.73   &22.96  &29.5  &1.06    &0.172\\
                &-3.5   &0.59   &10.27  &16.4  &0.73    &0.45
\end{tabular}
\end{ruledtabular}
 \footnotetext[1]{Measured at T=1.45 K}
\footnotetext[2]{The value of the Drude conductivity has been
obtained as described in Ref.~\onlinecite{our2}}
\footnotetext[3]{The parameters of the sample \#2 were analogous}
\end{table}

\begin{figure}
\center\includegraphics[width=12cm,clip=true]{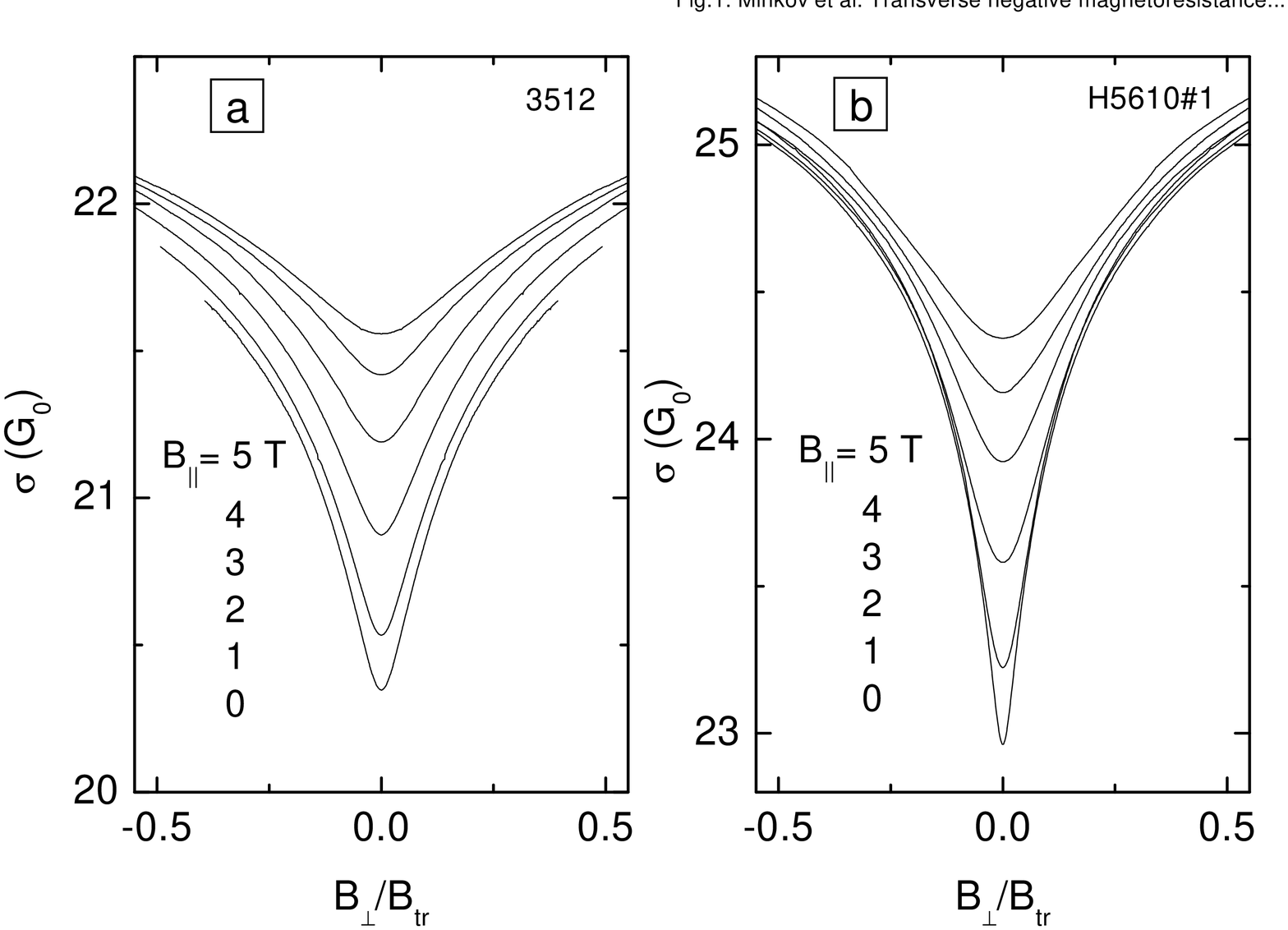} \caption{
The conductivity $\sigma$ as a function of $B_\perp$ dependences
measured at $T=1.45$~K for different in-plane magnetic fields for
structure 3512, $V_g=-1.5$~V (a) and structure H5610\#1,
$V_g=-2.5$~V (b). }
 \label{f1}
\end{figure}

\section{Results and discussion}
\label{sec:resultsWL} To make evident the difference in the effect
of in-plane magnetic field on the shape of negative
magnetoresistance at perpendicular field for these structures we
have presented  the data for both structures in Fig.~\ref{f1}
together. The magnetic field scale has been normalized to
characteristic for weak localization field
$B_{tr}=\hbar/(2el_p^2)$. One can see that in-plane magnetic field
changes the shape of magnetoresistance curve within wide range of
perpendicular field for structure 3512, while for structure H5610
the changes of the shape  occur at low perpendicular field only.
Below we demonstrate that this difference results from the
different correlation lengths of roughness in these structures.

\subsection{The role of short-range correlated roughness}
\label{ssec:3512} 
\begin{figure}
\center\includegraphics[width=10cm,clip=true]{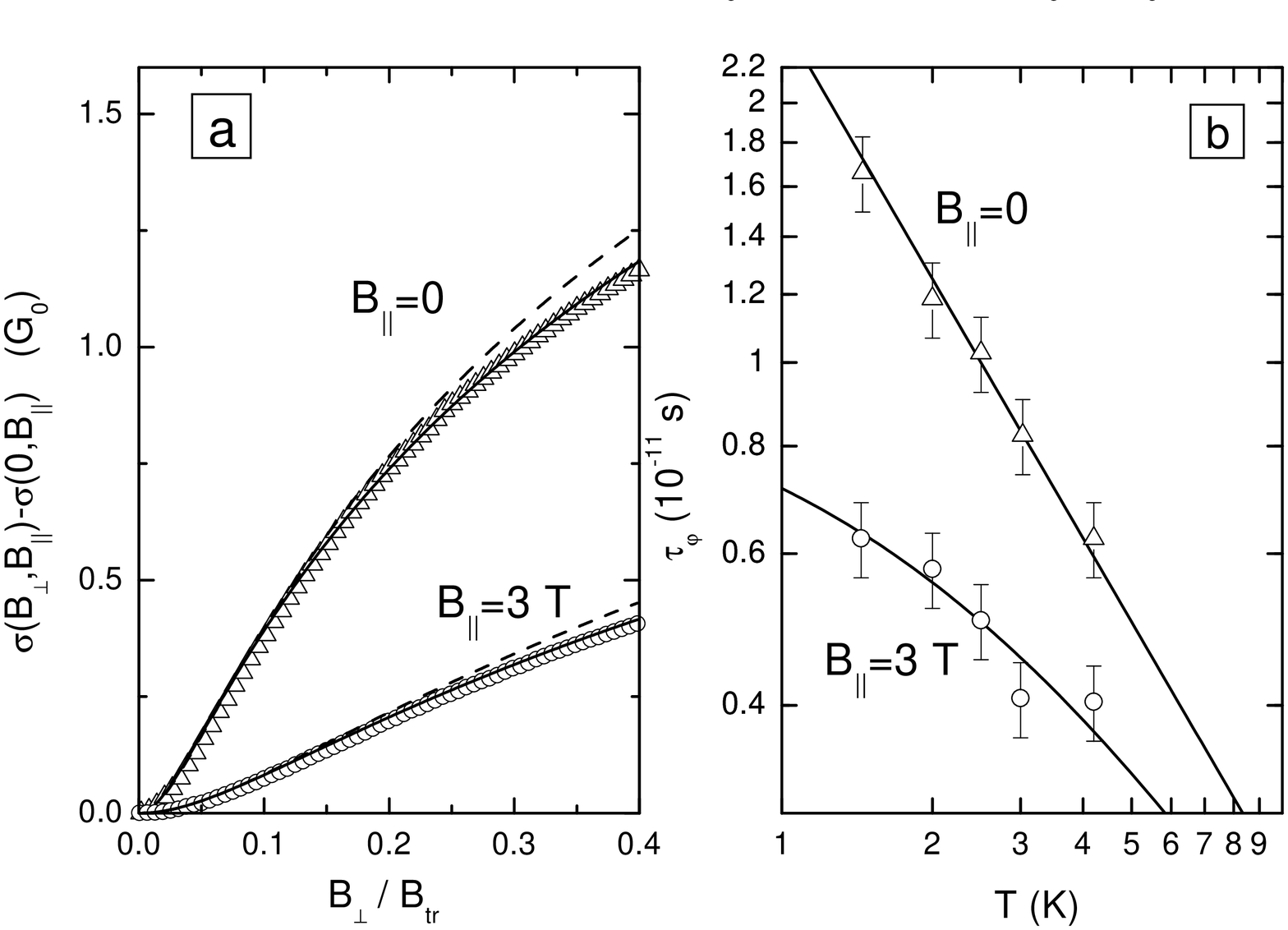} \caption{(a)
The
$[\sigma(B_\perp,B_\parallel)-\sigma(0,B_\parallel)]$-versus-$B_\perp$
dependences for structure 3512 at $B_\parallel=0$ and $3$~T,
$T=1.45$~K, $V_g=-1$~V. Symbols are the experimental data. Curves
are the best fit by Eq.~(\ref{eq01}) with parameters:
$B_\parallel=0$ -- $\alpha=0.98$ and $\tau_{\varphi}=1.5\times
10^{-11}$ s (dashed curve), $\alpha=0.87$ and
$\tau_{\varphi}=1.65\times 10^{-11}$ s (solid curve);
$B_\parallel=3$~T -- $\alpha=0.75$ and
$\tau_\varphi^\star=0.56\times 10^{-11}$ s (dashed curve),
$\alpha=0.62$ and $\tau_\varphi^\star=0.63\times 10^{-11}$ s
(solid curve). Dashed and solid curves correspond to the fitting
interval $B=(0-0.1)B_{tr}$ and $B=(0-0.2)B_{tr}$, respectively.
(b) The temperature dependence of the dephasing time for
$B_\parallel=0$ and $3$~T for structure 3512. Symbols are the
experimental data. Upper line is $T^{-1}$-law, lower one is drown
as described in the text.} \label{f2}
\end{figure}

Firstly, let us consider the data for the
structure 3512. Thorough studying of the weak localization
correction at $B_\parallel=0$ shows that experimental data are in
excellent agreement with conventional theory:

\begin{enumerate}
\item Magnetoconductance is well described by standard
Hikami-Larkin-Nagaoka (HLN) expression\cite{HLN}
\begin{eqnarray}
\Delta\sigma(B)&=&\rho_{xx}^{-1}(B)-\rho_{xx}^{-1}(0)=\alpha G_0 H(B,\tau_\varphi),\nonumber \\
H(B,\tau_\varphi) &\equiv & \psi\left(\frac{1}{2}+\frac{%
\tau_p}{\tau_\varphi}\frac{B_{tr}}{B}\right)
- \psi\left(\frac{1}{2}+\frac{%
B_{tr}}{B}\right)- \ln{\left(\frac{\tau_p}{\tau_\varphi}\right)}
\label{eq01}
\end{eqnarray}
with $\alpha$ and $\tau_\varphi$ as fitting parameters. In
Eq.~(\ref{eq01}), $G_0=e^2/(2\pi^2\hbar)$, $\psi(x)$ is a digamma
function, $\tau_\varphi$ is the phase breaking time. For strictly
diffusion regime ($\tau_p/\tau_\varphi\ll 1$, $B/B_{tr}\ll 1$) the
prefactor $\alpha$ has to be equal to unity. As Fig.~\ref{f2}(a)
illustrates, the values of the fitting parameters $\alpha$ and
$\tau_\varphi$ only slightly depend on the magnetic filed interval
in which the fit is done.

\item The temperature dependence of $\tau_\varphi$ is close to
$T^{-1}$-law [see Fig.~\ref{f2}(b)].

\item The temperature dependence of the conductivity at $B=0$ is
logarithmic. Slope of the experimental $\sigma/G_0$-versus-$\ln T$
dependence is about $1.45\pm0.05$. Here, the unity comes from the
weak localization and 0.45 comes from the electron-electron
interaction.\cite{our2}

\end{enumerate}

\begin{figure}
\center\includegraphics[width=13cm,clip=true]{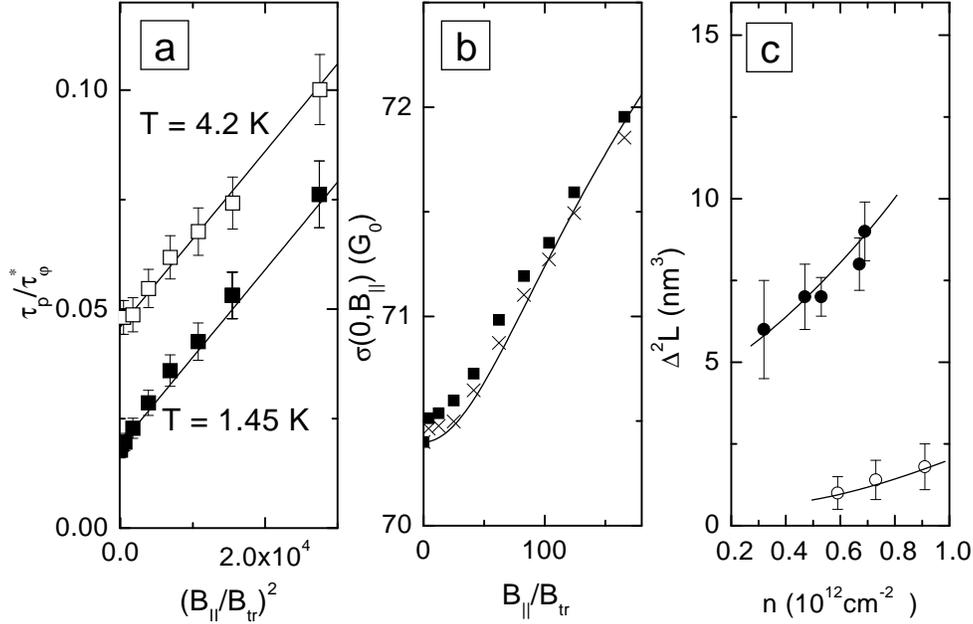} \caption{(a)
The value of $\tau_p/\tau^\star_\varphi$ as a function of
$B_\parallel^2$ for structure 3512 at $T=1.45$ and $4.2$~K,
$V_g=-1$~V. Symbols are experimental results. Lines are calculated
from Eqs.~(\ref{eq02}) and (\ref{eq03}) using
$\Delta^2L=7.2$~nm$^3$, $l_p=117$~nm, $\tau_p/\tau_\varphi=0.018$
($T=1.45$~K) and $0.048$ ($T=4.2$~K). (b) The conductivity as a
function of in-plane magnetic field for structure 3512, $T=1.45$,
$V_g=-1$~V. Squares are direct experimental measurements. Crosses
are obtained as
$\sigma(B_\parallel)=\sigma(B_\parallel=0)+\ln(\tau_\varphi/\tau_\varphi^\star)$,
where $\tau_\varphi$ and $\tau_\varphi^\star$ have been obtained
from the fit of experimental curve $\Delta\sigma(B_\perp)$ at
$B_\parallel=0$ and $B_\parallel\neq 0$, respectively. Solid line
is Eq.~(\ref{eq31b}) with $\Delta^2 L=7.2$~nm$^3$ and
$l_p=117$~nm. (c) The electron density dependence of the parameter
$\Delta^2 L$ for structures 3512 (solid symbols) and H5610 (open
symbols). Lines are provided as a guide to the eye.} \label{f3}
\end{figure}

Now let us analyze the data when in-plane magnetic field is
applied. As seen from Fig.~\ref{f2}(a) in this case the
magnetoconductance
$\sigma(B_\perp,B_\parallel)-\sigma(0,B_\parallel)$ is well
described by Eq.~(\ref{eq01}) also and the fitting parameters
$\alpha$ and $\tau_\varphi^\star$ (hereinafter, $\tau_\varphi$
relating to $B_\parallel\neq 0$ will be labelled as
$\tau_\varphi^\star$ ) depends only slightly on the fitting
interval. As clearly seen from Fig.~\ref{f3}(a)  the value of
$\tau_\varphi^\star$ strongly decreases when $B_\parallel$
increases.

The effect of in-plane magnetic field on the negative
magnetoresistance can be understand as follows. The weak
localization correction to the conductivity results from the
interference of electron waves scattered along closed trajectories
in opposite directions (time-reversed paths). Magnetic field gives
the phase difference between pairs of time-reversed paths
$\varphi=2\pi{\bf B S}/\Phi_0$ (where $\Phi_0$ is the quantum of
magnetic flux, $\bf S$ is the algebraic area enclosed) and thus
destroys the interference and results in negative
magnetoresistance. The scalar product $\bf BS$ is zero for ideal
2D structures at in-plane magnetic field therefore this field does
not destroy the interference, and the negative magnetoresistance
is absent in this magnetic field orientation. In real 2D structure
the mean electron position in growth direction randomly changes at
motion along closed paths due to interface roughness. Therefore
the product $\bf BS$ becomes nonzero for in-plane magnetic field
that leads to suppression of the weak localization correction.
Theoretical analysis \cite{Wheeler,Mathur,Malshukov}shows that for
the case of short-range correlated roughness the role of in-plane
magnetic field reduces to increasing of the dephasing rate
\begin{equation}
{1 \over \tau^\star_\varphi}= {1 \over \tau_\varphi}+ {1 \over
\tau_\parallel} \label{eq02}
\end{equation}
where $\tau_\parallel^{-1}$  is determined by parameters of
roughness \cite{Mathur}
\begin{equation}
{1 \over \tau_\parallel}\simeq {1 \over \tau_p} {\sqrt{\pi}\over 4
} {\Delta^2 L \over l_p^3}\left({B_\parallel \over
B_{tr}}\right)^2\, . \label{eq03}
\end{equation}
Here, $\Delta$ is the root-mean-square height of the fluctuations,
and $L$ is the distance over which the fluctuations are
correlated.

Let us consider how our experimental results for structure 3512
agree with this model. Fig.~\ref{f3}(a) shows that
$\tau_p/\tau^\star_\varphi$ increases linearly with
$B_\parallel^2$ in a full agreement with Eqs.~(\ref{eq02}) and
(\ref{eq03}), therewith the slope of this dependence is
temperature independent. In framework of this model the
temperature dependence of $\tau^\star_\varphi$ at the presence of
in-plane magnetic field has to saturate on the value
$\tau_\parallel$ with decreasing temperature. Figure~\ref{f2}(b)
in which the experimental results obtained for $B_{\parallel}=3$~T
are plotted shows that $\tau^\star_\varphi$-versus-$T$ dependence
really tends to saturate at $T\to 0$. In the same figure we plot
the $\tau^\star_\varphi$-versus-$T$ curve calculated in accordance
with Eq.~(\ref{eq02}). In this calculation the dependence
$2.5\times 10^{-11}/T$ which is a good interpolation of
experimental data for $B_\parallel=0$ [see Fig.~\ref{f2}(b)] has
been used as $\tau_\varphi(T)$ in right-hand side of
Eq.~(\ref{eq02}). The quantity $\tau_\parallel^{-1}=1\times
10^{11}$~s$^{-1}$ has been obtained as a difference between two
values $(\tau^\star_\varphi)^{-1}$ and $\tau_\varphi^{-1}$ found
experimentally at $T=1.45$~K. Good agreement is evident within
whole temperature range.

This model predicts also that growth of in-plane magnetic field
has to lead to increase of conductivity at $B_{\perp}=0$ as
follows
\begin{subequations}
\label{eq31}
\begin{eqnarray}
\sigma(0,B_\parallel)&=&\sigma(0,0)+G_0\ln{\tau_{\varphi} \over
\tau^\star_\varphi} \label{eq31a} \\
&\simeq &\sigma(0,0)+G_0\ln\left[1+{\tau_\varphi \over \tau_p}
{\sqrt{\pi}\over 4 } {\Delta^2 L \over l_p^3}\left({B_\parallel
\over B_{tr}}\right)^2\right]\, . \label{eq31b}
\end{eqnarray}
\end{subequations}
In Fig.~\ref{f3}(c) we present the in-plane magnetic field
dependence of the conductivity which was measured directly and was
calculated from Eq.~(\ref{eq31a}) using $\tau^\star_\varphi$ found
above [see Fig.~\ref{f3}(a)]. One can see that within experimental
error these data agree each other satisfactorily. Thus, all
effects predicted for the case of the short-range correlated
roughness are observed in the structure 3512. Therefore we believe
that the slope of
$\tau_p/\tau^\star_\varphi$-versus-$(B/B_{tr})^2$ dependence gives
the parameter of roughness $\Delta^2L$ which with the use of
$l_p=117$~nm  for $V_g=-1$~V can be estimated as $\Delta^2L\simeq
7.2$ nm$^{3}$. Naturally, Eq.~(\ref{eq31b}) with this value of
$\Delta^2L$ well describes the experimental in-plane magnetic
field dependences of the conductivity, measured without
perpendicular magnetic field [Fig.~\ref{f3}(b)].

We have carried out such analysis for various gate voltages and
plotted the electron density dependence of $\Delta^2L$ in
Fig.~\ref{f3}(c). One can see that the value of $\Delta^2L$
somewhat decreases with decreasing electron density. The following
model can explain this observation. The inner interface lying in
the depth of the structure has smaller roughness than that lying
closer to the cap layer. With the decrease of the gate voltage,
i.e.,  with decrease of the electron density, the wave function is
hold closer against the inner interface resulting in the reduction
of the role of the more rough outer interface. The larger
roughness of the outer interface seems to be natural for the
quantum well heterostructures studied because lattice mismatch
leads to a strain of In$_x$Ga$_{1-x}$As layer and corrugation of
outer interface. Analogous results were obtained in
Ref.~\onlinecite{Wheeler} for silicon MOSFET.

\subsection{Effect of nanoclusters on the weak localization}
\label{ssec:H5610}

\begin{figure}
\center\includegraphics[width=10cm,clip=true]{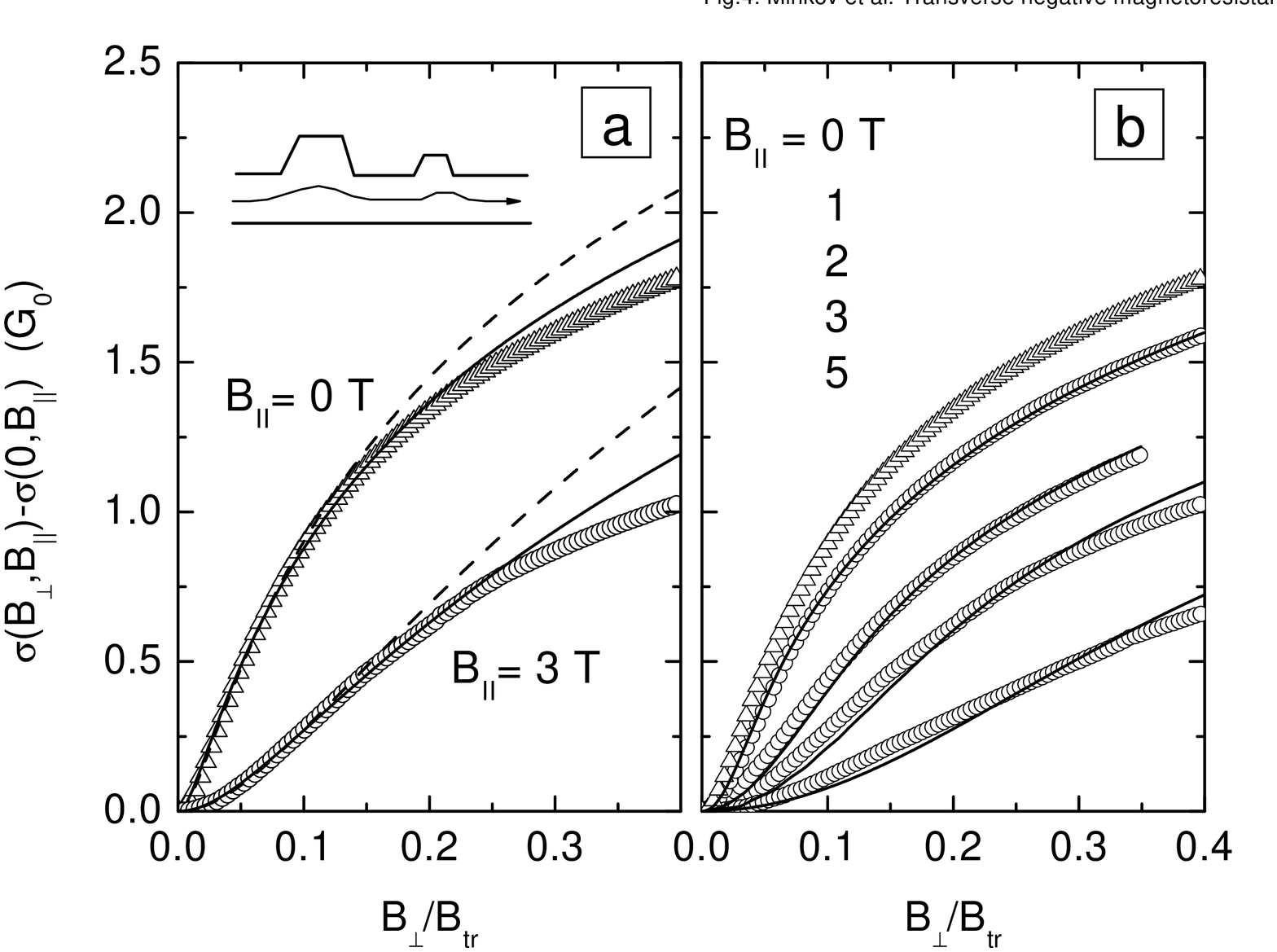} \caption{
The
[$\sigma(B_\perp,B_\parallel)-\sigma(0,B_\parallel)$]-versus-$B_\perp$
dependences for structure H5610\#2 taken  at  $T=1.45$~K and
$V_g=-2.5$~V. Symbols are the experimental data. Curves in (a) are
the best fit by Eq.~(\ref{eq01}) with parameters: $B_\parallel=0$
-- $\alpha=1.0$, $\tau_\varphi=1.2\times 10^{-11}$ s (dashed line)
and $\alpha=0.9$, $\tau_\varphi=1.45\times 10^{-11}$~s (solid
line); $B_\parallel=3$~T -- $\alpha=2.2$, $\tau_\varphi=2.3\times
10^{-12}$ s (dashed line) and $\alpha=1.4$,
$\tau_\varphi=2.9\times 10^{-12}$~s (solid line). Dashed and solid
curves correspond to the fitting interval $B=(0-0.1)B_{tr}$ and
$B=(0-0.2)B_{tr}$, respectively. Curves in (b) are the best fit by
Eq.~(\ref{eq04}) in which the Gaussian distribution and
experimental curve $\sigma(B_\perp,0)$ are used for $F(\beta)$ and
$\delta\sigma(B_\perp+\beta B_\parallel,\tau_\varphi)$,
respectively. The values of fitting parameter $\Delta_\beta$
sequence from $B_\parallel=1$~T to $B_\parallel=5$~T are the in
following: 0.34; 0.41; 0.47; 0.52 degrees. Inset in (a) is a
schematic representation of electron motion along the quantum well
with one rough side.} \label{f4}
\end{figure}

Now let us consider the effect of in-plane magnetic field on the
negative magnetoresistance for structure H5610 with nanoclusters.
The
[$\sigma(B_\perp,B_\parallel)-\sigma(0,B_\parallel)$]-versus-$B_\perp$
plot for structure H5610\#2 is presented in Fig.\ref{f4}. As seen
from  Fig.\ref{f4}(a) the negative magnetoresistance measured at
$B_{\parallel}=0$ is perfectly described by Eq.~(\ref{eq01}). If
one tries to fit by Eq.~(\ref{eq01}) the data measured at
$B_{\parallel}\neq0$ one finds that the fitting parameters
strongly depend on the fitting interval. For instance, the
prefactor $\alpha$ strongly decreases from $\alpha=2.2$ to
$\alpha=1.4$ when the fitting interval of $B$ is expanded from
$0-0.1$ to $0-0.2$  [see Fig.\ref{f4}(a)]. Moreover, the
significantly higher than unity value of the prefactor is
unreasonable for single quantum well structure with one occupied
subband. All this points to the fact that Eq.~(\ref{eq01}) rather
poorly describes the experimental data for structure H5610. We
believe that such a behavior is sequence of a long-range
correlated roughness which is caused by nanoclusrers. The presence
of nanoclusters in this structure leads to smooth random deviation
in position of an electron in growth direction when it moves over
the quantum well [see inset in Fig.~\ref{f4}(a)]. Influence of
in-plane magnetic field on the shape of magnetoresistance curve in
perpendicular field for the case $l_\varphi>L>l_p$  was
theoretically studied in Ref.~\onlinecite{Mathur}. However the
final expressions are very complicate and cumbersome to compare
them with the experimental curves directly.

Another limiting case $L>l_\varphi$ is very simple and transparent
from the physical point of view. In this case one can consider
that all the actual closed paths lie on the flat elements of size
larger than $l_\varphi$, which are inclined from ideal 2D plane
through small random angles $\beta$. This means that the resulting
magnetoresistance is a sum of the contributions of these deflected
elements. The contribution of each element is
$\delta\sigma(B_n,\tau_\varphi)=\delta\sigma(B_\perp+\beta
B_\parallel,\tau_\varphi)$, where $B_n$ is projection of the total
magnetic field onto the normal to the element plane. Then, the
total magnetic field dependence of the conductivity can be written
as
\begin{equation}
\sigma(B_\perp,B_\parallel,\tau_\varphi)=\int d  \beta
F(\beta)\delta\sigma(B_\perp+\beta B_\parallel,\tau_\varphi),
 \label{eq04}
\end{equation}
where  $F(\beta)$ is the distribution function of the deflection
angles. To compare this expression with experimental data one
needs to specify the functions in the right-hand side of
Eq.~(\ref{eq04}). We have used the Gaussian distribution for
$F(\beta)$ with root-mean square $\Delta_\beta$. The experimental
$\sigma$-versus-$B_\perp$ curve measured at $B_\parallel=0$  has
been used as $\delta\sigma(B_\perp+\beta
B_\parallel,\tau_\varphi)$.  The result of the fitting procedure
for $\sigma(B_\perp,B_\parallel)-\sigma(0,B_\parallel)$ with one
fitting parameter $\Delta_\beta$ is shown in Fig.~\ref{f4}(b). One
can see that this simple model perfectly describes the shape of
the experimental magnetoresistance curve in the presence of
in-plane magnetic field up to $B_\parallel=2$~T and the parameter
$\Delta_\beta$ found from the fit is really small in magnitude:
$\Delta_\beta\simeq 0.3^\circ-0.4^\circ$. A noticeable discrepancy
between this model and experimental observations is evident at
higher magnetic field, $B\gtrsim 3$~T, the parameter
$\Delta_\beta$ sufficiently increases with $B_\parallel$ increase
[see Fig.~\ref{f6}(a)]. To our opinion, the situation when
$\Delta_\beta$ is independent of in-plane magnetic field seems
more natural.

\begin{figure}
\center\includegraphics[width=10cm,clip=true]{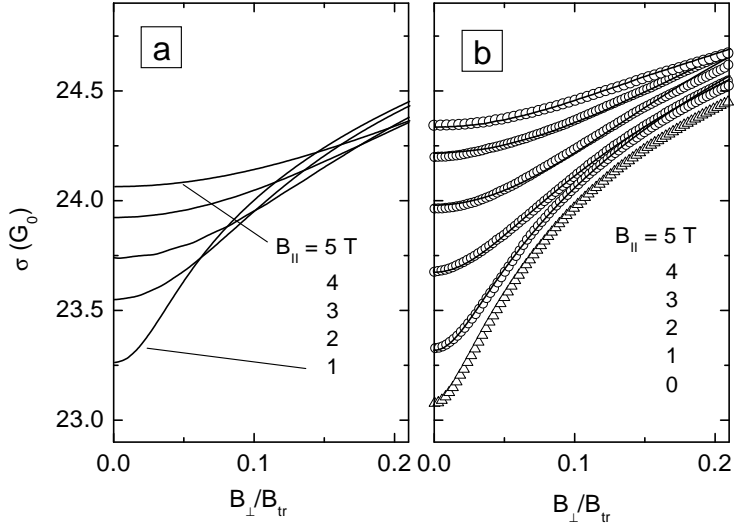} \caption{(a)
The $B_\perp$-dependences of absolute values of the conductivity
taken at different in-plane magnetic field. Lines in (a) are the
just the same as in Fig.~\ref{f4}(b) but plotted without
subtraction of the value of $\sigma(0,B_\parallel)$. Symbols in
(b) are the experimental data, lines are obtained taking into
account both long- and short-range correlated roughness.  The
fitting parameters as a functions of in-plane magnetic field are
shown in Fig.~\ref{f6}. } \label{f5}
\end{figure}

Such a discrepancy between this model and experimental
observations evident at $B_\parallel\gtrsim 3$~T can be understood
if one supposes a simultaneous existence of short- and long-range
correlated roughness in the structure H5610. As shown above the
short-range correlated roughness results effectively in lowering
of $\tau_\varphi$ in in-plane magnetic field. Thus, it becomes
meaningless to use the experimental $\sigma$-versus-$B_\perp$
curve measured at $B_\parallel=0$ in right-hand side of
Eq.~(\ref{eq04}) when $B_\parallel$ is rather high.

The presence of short-range correlated roughness in structure
H5610 is more pronounced when considering the effect of in-plane
magnetic field on the absolute value of the conductivity.
Figure~\ref{f5}(a) shows the curves shown in Fig.~\ref{f4}(b) but
plotted without subtraction of $\sigma(0,B_\parallel)$. Comparing
this figure with Fig.~\ref{f5}(b), in which the experimental
results are presented, one can see that the model taking into
consideration only the long-range correlated roughness does not
describe the behavior of absolute value of $\sigma$ in in-plane
magnetic field. It is most conspicuous at $B_\perp/B_{tr}\gtrsim
0.1$, where the experimental $\sigma$-versus-$B_\perp$ plots are
shifted up with $B_\parallel$-increase whereas the calculated
curves  tend to merge together. It is natural to suggest that the
shift of experimental curves is a result of the influence of short
range correlated roughness which leads to decrease of
$\tau_\varphi$ and, thus, to increase of the conductivity with
increasing of in-plane magnetic field when the perpendicular field
is fixed.

To take into account coexistence of both long- and short-range
correlated roughness in our model, we have used the quantity
$\sigma(B_\perp=0,B_\parallel=0)+\alpha G_0 H(B,\tau_\varphi)$ as
$\delta\sigma(B_\perp+\beta B_\parallel,\tau_\varphi)$ in
Eg.~(\ref{eq04}), where $\sigma(B_\perp=0,B_\parallel=0)$ is
measured experimentally. Thus, we can manipulate by three fitting
parameters $\alpha$, $\tau_\varphi$, and $\Delta_\beta$ to
describe quantitatively the experimental results for structure
H5610. Figure~\ref{f5}(b) illustrates excellent agreement between
experimental results and the model taking into account both types
of roughness.

Let us now consider whether the fitting parameters are reasonable.
First of all, the value of the prefactor is about $0.8-0.9$ that
agrees with sufficiently large $\tau_\varphi$ to $\tau$ ratio at
any $B_\parallel$: $\tau_\varphi/\tau\simeq 100-200$. Second, the
parameter $\Delta_\beta$ does not practically depend on
$B_{\parallel}$ and is close to that obtained at $B_\parallel<3$~T
without taking into account the short-range correlated roughness
[see Fig.~\ref{f6}(a)]. Thus, the value of $\Delta_\beta$
characterizing the long-range correlated roughness can be
estimated at $0.35^\circ$. Finally, the fitting parameter
$1/\tau_\varphi^\star$ exhibits quadratical increase when
$B_\parallel$ increases [see Fig.~\ref{f6}(b)], that allows us to
estimate the scale of the short-range correlated roughness using
Eqs.~(\ref{eq02}) and (\ref{eq03}). The value of $\Delta^2 L$ in
structure H5610 with nanoclusters occurs to be about
$1.2\,\text{nm}^3$ that is less than that for structure 3512. Such
an analysis carried out for other gate voltages shows that the
parameter $\Delta_\beta$ is independent of electron density within
experimental error and, thus, is about $0.35^\circ$ and the
parameter $\Delta^2 L$ increases from approximately $1$ to
$1.8\,\text{nm}^3$ when the electron density varies from
$0.59\times 10^{12}$ to $0.91\times 10^{12}$~cm$^{-2}$ [see
Fig.~\ref{f3}(c)]. As for structure 3512 (see Section
\ref{ssec:3512}), we believe that such a behavior of $\Delta^2L$
is a result of the shift of the wave function to inner smooth
interface of the quantum well that in its turn leads to reduction
of the role of outer rough interface.
\begin{figure}
\center\includegraphics[width=10cm,clip=true]{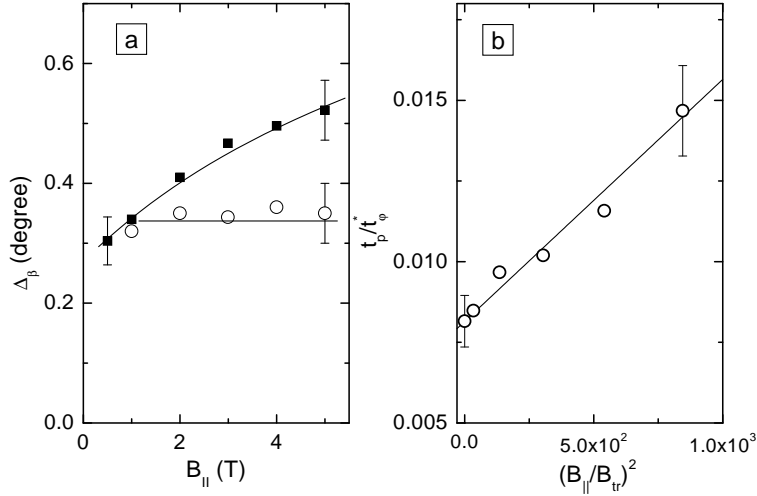} \caption{The
fitting parameters $\Delta_\beta$ (a) and $\Delta^2L$ (b)
corresponding to solid lines in Fig.~\ref{f5}(b) as functions of
in-plane magnetic field. Solid symbols correspond to the
long-range roughness model, open symbols are obtained when both
short- and long-range roughness are taken into consideration.
Lines in (a) are provided as a guide to the eye, line in (b) is
calculated from Eqs.~(\ref{eq02}) and (\ref{eq03}) using
$\Delta^2L=1.4$~nm$^3$ and experimental values $l_p=44$~nm and
$\tau_p/\tau_\varphi=8.15\times 10^{-3}$.} \label{f6}
\end{figure}

Thus, for the 2D structure with nanoclusters we can adequately
describe the influence of in-plane magnetic field on weak
localization combining two limiting theoretical models
corresponding to short- and long-range correlated roughness.

\subsection{Results of AFM-studies}
\label{ssec:AFM} 
\begin{figure}
\center\includegraphics[width=12cm,clip=true]{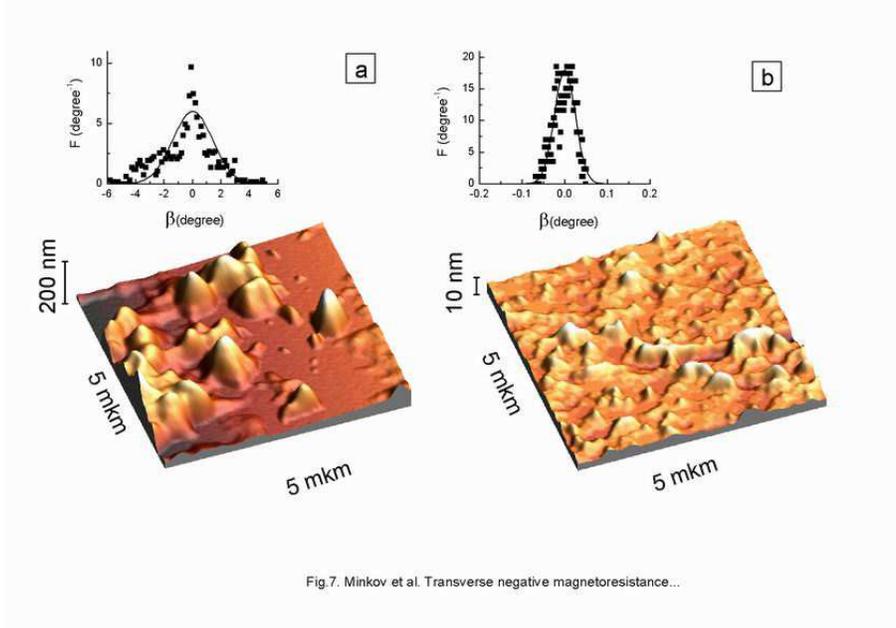} \caption{The
AFM images for structure H5610 (a) and 3512 (b) obtained after
removing of the cap layer. Insets show the angle distribution
function $F(\beta)$ obtained for ${\cal L}=2l_\varphi$ (see text)
The values of $l_\varphi$ at $T=1.5 K$ are $490$ nm and $870$ nm
for structures H5610 and 3512, respectively. } \label{f7}
\end{figure}

To assure that the structure H5610 distinguishes
from the structure 3512 by the presence of long-range correlated
roughness and to estimate its parameters, we have attempted to
measure the profile of the quantum well surface. For this purpose
the cap layer was removed using the selective
etching.\cite{selEtch} After that the surface was scanned by
Atomic Force Microscope (AFM) using TopoMetrix Accurex TMX-2100
ambient air AFM in Contact Mode. Si$_3$N$_4$ pyramidal probes were
employed. The AFM-images for both structures are shown in
Fig.~\ref{f7}. It is clearly seen that the scales of surface
roughness are drastically different.
In order to get the quantitative information corresponding to our
experiments, we have processed the images.

Let us firstly consider the long-range roughness. In accord with
the model used for interpretation of the results for structure
H5610 the surfaces presented in Fig.~\ref{f7} were approximated by
set of the flat squares  of size ${\cal L}>l_\varphi$, then the
angle distribution function $F(\beta)$ entering in
Eq.~(\ref{eq04}) was found. This function is presented in insets
in Fig.~\ref{f7} for both structures. It has been approximated by
the Gaussian distribution and  the dispersion $\Delta_\beta$ has
been found. The value of $\Delta_\beta$ obtained for ${\cal
L}=2l_\varphi$ and ${\cal L}=3l_\varphi$ was close and differed by
30\%.

For structure H5610 the value of $\Delta_\beta$ is about $2^\circ$
that is five-six times larger than the dispersion obtained from
the weak localization measurements [see Fig.~\ref{f6}]. The reason
for such a discrepancy is qualitatively clear. In reality, an
electron moves not over the surface, it moves in the quantum well
laying under the surface. Therefore, the deviations of electron in
$z$-direction are smaller than the roughness magnitude. Thus, we
consider the results of weak localization and AFM experiments
being in a satisfactory agreement for structure H5610.

For structure 3512, the dispersion is about $0.035^\circ$, i.e.,
the long range correlated roughness is practically absent. This
fact agrees with the experimental result that only the short range
correlated roughness reveals itself in weak localization in the
presence of in-plane magnetic field.

\begin{figure}
\center\includegraphics[width=12cm,clip=true]{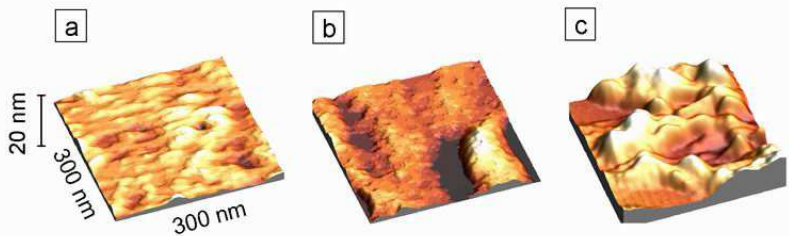} \caption{The
AFM images for structure 3512 (a) and H5610 (b,c), scanned area is
$300\times 300$~nm.} \label{f8}
\end{figure}

To estimate the parameter $\Delta^2 L$ responsible for the
influence of the short-range correlations on the weak
localization, the surfaces have been scanned with the higher
resolution (Fig.~\ref{f8}). The mean peak spacing for structure
3512 found from AFM-image given in Fig.~\ref{f8}(a) is $65$~nm,
that is really less than mean free path ($l_p=210$~nm for
$V_g=0$). So the use of the short-range correlated roughness model
is justified. The value of $\Delta$  found as a root-mean square
deviation in $z$-direction is about $0.35$~nm. So, we obtain from
independent AFM-measurements the parameter $\Delta^2 L\simeq
8$~nm$^3$ which is close to that obtained from weak-localization
experiment [see Fig.~\ref{f3}(c)].

It is more difficult to carry out the analogous estimation for
structure H5610. First, the flat and hilly areas look differently
at this resolution [see Figs.~\ref{f8}(b) and \ref{f8}(c)].
Second, the peak spacing distribution is very wide and has no
maximum.  Nevertheless we try to estimate $\Delta$ using $l_p/2$
as $L$.  The value of $\Delta$ found for the flat and hilly areas
appears to be different: $0.2$ and $0.6$ nm, respectively.
Therefore, the value of parameter $\Delta^2 L$ for these areas are
significantly different $1.6$~nm$^3$ and $14$~nm$^3$ (we used here
$l_p=80$~nm that corresponded $V_g=0$). Recall that $\Delta^2 L$
obtained from the weak localization experiments is $(2\pm 0.5)$~nm
[see Fig.~\ref{f3}(c)]. Taking into account the large scatter of
AFM results, we consider such an agreement as satisfactory.

\section{Conclusion}
\label{sec:concl} We have experimentally studied the effects of
in-plane magnetic field on the interference induced negative
magnetoresistance in perpendicular magnetic field for different
types of quantum well heterostructures. It has been shown that the
effects significantly depend on the relationship between mean free
path and in-plane size of the roughness. The analysis of the shape
of the negative magnetoresistance at in-plane magnetic field gives
possibility to recognize the characteristic in-plane scale of the
roughness and estimate its parameters. The results of weak
localization studies have been found in a good agreement with
evidence from AFM measurements.

\section*{Acknowledgments}

This work was supported in part by the RFBR through Grants No.
01-02-17003, No. 01-02-16441, No. 03-02-16150, and
No.~03-02-06025, the CRDF through Grants No. REC-001 and No.
REC-005, the Program {\it University of Russia} through Grant No.
UR.06.01.002, and the Russian Program {\it Physics of Solid State
Nanostructures}.

\end{document}